# AI and Structural Injustice

## Foundations for Equity, Values, and Responsibility

**Johannes Himmelreich**, Department of Public Administration and International Affairs, Maxwell School at Syracuse University, jrhimmel@syr.edu

**Désirée Lim**, Department of Philosophy at the Pennsylvania State University, dyl10@psu.edu

**Abstract**
This chapter argues for a structural injustice approach to the governance of AI. Structural injustice has an analytical and an evaluative component. The analytical component consists of structural explanations that are well-known in the social sciences. The evaluative component is a theory of justice. Structural injustice is a powerful conceptual tool that allows researchers and practitioners to identify, articulate, and perhaps even anticipate, AI biases. The chapter begins with an example of racial bias in AI that arises from structural injustice. The chapter then presents the concept of structural injustice as introduced by the philosopher Iris Marion Young. The chapter moreover argues that structural injustice is well suited as an approach to the governance of AI and compares this approach to alternative approaches that start from analyses of harms and benefits or from value statements. The chapter suggests that structural injustice provides methodological and normative foundations for the values and concerns of Diversity, Equity, and Inclusion. The chapter closes with an outlook onto the idea of "structure" and on responsibility. The idea of a structure is central to justice. An open theoretical research question is to what extent AI is itself part of the structure of society. Finally, the practice of responsibility is central to structural injustice. Even if they cannot be held responsible for the existence of structural injustice, every individual and every organization has some responsibility to address structural injustice going forward.

# 1. Introduction

Structural injustice has risen to the top of the agenda. The United States continues to be rocked by police-inflicted violence, especially against people of color; in the background, the American foundational legend—of having discovered a "new world" on which generations of settlers built their dreams—is falling apart.[1] At the same time, countries in Europe are grappling with disparities in how gender, race, ability, class, and origin affect wages, social recognition, political and economic participation, and casually-inflicted prejudices that perfuse everyday life.[2] Around the globe, the pervasive effects of racism and colonialism have been documented: not only in terms of material differences in outcomes and opportunities, but also in discursive or semiotic differences in how particular groups are portrayed. For example, Black neighborhoods in New York City are advertised as exotic and edgy destinations for the adventurous white tourist (Törnberg & Chiappini, 2020). In short, the Global North is growing the awareness that the *status quo* builds on and manifests various past and present-day injustices and normative deficiencies.

---

[1] This legend may never have been plausible outside the lived experience of a middle- to upper-class white population.

With this growing awareness, students, professionals, academics, and policy-makers in the Global North are attending to the role that Artificial Intelligence (AI) plays in maintaining, entrenching, or even exacerbating this unjust *status quo*.[3] A software feature may be well-intended and an algorithm considered objective, but when deployed in an unjust *status quo*, they will likely perpetuate injustice or worsen it. AI interacts with unjust social structures—AI exacerbates structural injustice.

What should we make of this idea? The term "structural injustice" has been used to describe a very wide range of different phenomena, including police brutality, unequal health outcomes, pay gaps, as well as inequalities in educational opportunities. But the theory of structural injustice faces challenges. The theory is epistemically hard. Even if you know there *is* a problem of structural injustice, you may find understanding what this "problem" is exactly just as hard as solving it. Moreover, the theory of structural injustice might seem methodologically deficient. Insofar as it is not couched in terms of agents or institutions, or doesn't specify causal mechanisms, skeptics decry the idea of structural injustice as a set of, perhaps ideologically-motivated, causally free-flowing conjectures. The theory of structural injustice is also rhetorically or ideologically disadvantaged. Concerns about structural injustice are often articulated in opposition to a liberal or western mainstream, sometimes as issues of diversity, equity, and inclusion (DEI). But what, exactly, lies behind this language of DEI?

This chapter presents the theory of structural injustice. We explain some of the theoretical foundations of equity and social justice, or, relatedly DEI. We demonstrate that, as opposed to being a mere buzzword, this theory is not a political slogan but a respectable normative and empirical concept. As such, structural injustice ought to inform research and legislation on issues, such as, how AI interacts with social identities like gender and race, or how the development of AI reflects existing economic interests. We argue that structural injustice, and AI's capacity to exacerbate it across many spheres of human life, must be taken seriously.

We concentrate here mostly on gender and race. Structural injustices usually attach to salient social categories such as race, gender, age, ability, or sexual orientation. As such, structural injustice has to do with identity. Because race, gender, age, ability, or sexual orientation are central to a person's self-conception or self-image, how one is treated based on these characteristics is a matter of high moral concern and deep emotional valence. But although structural injustice attaches to social identity, this chapter is not concerned with identity politics.[4] Insofar as the chapter aims to make a political contribution at all, ours has a theoretical focus: We are concerned that appeals to the values of DEI or to structural injustice in AI are dismissed as a fad, as mysterious methodology, as empty slogans used to virtue-signal, or as sectarian interests pushed by advocacy groups. This is not the case. We argue that the theory of structural injustice is a useful lens for anyone concerned with AI governance. A theory of structural injustice allows researchers and practitioners to identify, articulate and perhaps even anticipate phenomena and problems that may otherwise go unrecognized. Some basic theoretical ingredients of structural injustice—social-structural explanations and theories of justice—easily extend from gender and race to issues of ability, sexual identity, or economic power.

---

[2] Such "everyday" instantiations of prejudice are popularly known as "microaggressions". For a detailed philosophical account of microaggressions and the injustice they may perpetuate, see Rini (2020).
[3] We understand "AI" as statistical methods that use Machine Learning (ML) to make data-based predictions or decisions. With the increasing availability of data, and the decreasing costs of storage and computing, the possible uses of AI have dramatically increased.

This chapter covers only limited ground. First, it does not cover business or managerial issues, such as building project teams, nor does it cover the sociology, demographics, or politics of who builds and uses AI today. Making sure that research, design and engineering teams are diverse is important for addressing structural injustices and for bringing the values of DEI to life. This chapter concentrates instead on the foundations of DEI or social justice: Which theoretical and moral considerations underwrite a concern for DEI? Second, this chapter concentrates on stylized examples. We cannot do justice to the nuanced ways in which structural injustice plays out. This is because social identities are complex, and the experience of structural injustice often attaches to more than one social category at once. We take our examples mainly from a north-american context. However, thanks to them being somewhat abstract, the examples should easily travel to other contexts.

In sum, this chapter provides a conceptual lens to bring into sharper focus how AI relates to structural injustice. We contribute to discussions on DEI by conceptualizing present-day calls for diversity, equity, and inclusion in the sphere of AI as *demands of justice*, rather than a bid to reduce harm or attempts to comply with ethical codes of conduct that are frequently drafted by tech corporations. We do this by giving a primer on the concept of structural injustice as it has been developed by scholars in moral and political philosophy, in particular by Iris Marion Young (Young, 2006, 2011). We argue that this perspective on structural injustice is particularly important for AI. This chapter not so much highlights the myriad unintended ways in which developing and deploying AI may contribute to structural injustice, but it aims to equip the reader with the theoretical foundations for tools that help to recognize structural injustices and with a sense of responsibility to attend to them.

---

[4] Perhaps structural injustice should *not* be approached as a matter of identity at all, but as a matter of difference (Young, 2009).

# 2. AI and Structural Injustice

As a starting point, we assume that police violence, gender prejudice, and—more broadly—disadvantages because of differences in class, race, ability, or gender are often *structural phenomena*. By this, we mean, first, that they are generally not fully explained by the intentional actions of particular individuals.[5] Police officers may act out of basic instinctive self-preservation or because they internalize problematic informal professional norms, and journalists ask questions that are loaded with gender prejudices because they have reason to believe that viewers care about these questions. Similarly, even the measurable physiological differences—that African American women are 60 percent more likely to have high blood pressure, or that African American children have vastly increased Blood Lead Levels—cannot be linked to any individual act or policy at all. Second, structural phenomena—police violence and elevated lead levels—often seem to have something in common; not only in their outcomes (each concerns a disadvantage on Black people) but also in their causes.

Well-intentioned individuals—even those who care about the end-goal of racial equality—typically take social structures for granted and accept their constraints and affordances. We tend to work with our world as it is "given". Informal norms, unconsciously internalized expectations, or learned emotional responses are larger than individuals but have a hold on us. Such structural features influence and explain the conduct of individuals and, even more so, aggregate outcomes. Engineers, entrepreneurs, and policy-makers may put great effort behind building products that make the world a better place and that yet still may have the opposite effect. These effects may not be deliberate, but nevertheless, they are profound. To understand this, structural explanations can help. Structural explanations are the central ingredient in a theory of structural injustice (cf. Soon, 2021). Let us turn to an illustrative example.

## 2.1 How Structure Matters: Example of AI in Health Care

Hospitals use decision-rules, i.e. algorithms, to identify high-need patients—patients with chronic conditions and complex medical needs—to invite them to special outpatient treatment programs. In 2019, researchers found that an algorithm to identify such high-need patients exhibits "significant racial bias" (Obermeyer et al., 2019): Although more than 46% of Black patients should have received the help of such treatment programs, only 17.7% did. Black patients were less likely to be selected for the special treatment programs despite there being no difference in underlying health conditions.

The reason for this bias is instructive. On the face of it, from the perspective of researchers and policymakers, the algorithm seems unbiased. In fact, it predicts total medical expenditures without a significant racial difference. The algorithm assigns patients risk scores—this is the prediction it makes—and each of these risk scores is associated with the same average medical expenditures, regardless of whether a patient is white or Black. The algorithm hence correctly predicts, without significant bias, whether or not a patient has high health needs—at least when "high need" is understood as total medical expenditures.

---

[5] Intentional racism etc. is a problem—and a glaring one. What is wrong here is relatively easy to explain, in contrast to structural injustice.

The bias creeps in, however, when we consider the relationship between health and medical expenditures. When developing the algorithm at hand, medical expenditures were falsely taken to be a good proxy for underlying health. On average, however, the healthcare system spends *less* on Black patients than on white patients *at all levels of healthcare needs*. As a result, the algorithm *did* predict *costs* correctly, but given that average healthcare costs are lower for Black patients, the algorithm underestimated the Black patients' healthcare needs. In a way, the algorithm "sees" fewer health needs in Black patients. This lines up with how Black patients' experience their interactions with the medical profession already today (Cuevas et al., 2016). AI thus overlooks Black persons' healthcare needs, rendering them less visible or less urgent.

This is a clear case of racial *injustice* (we leave aside for now what makes it an injustice: why, since not all differences are unjust, is *this* difference an injustice?). This injustice is *structural* because it is best explained with reference to structural features. To get a sense for the power of structural explanations, consider why, for any given health need, average medical expenditures are lower for Black patients.

One set of hypotheses looks at medical professionals. Doctors need not be outright racists, in the sense of consciously ascribing racial inferiority to Black patients,[6] in order to treat them differently. We inhabit a world where racial categories remain socially salient: that is, our perceived racial difference has a significant impact on how our social interactions unfold across a wide range of contexts. Many people still believe that race is a biobehavioral essence that explains our behavioral and cultural dispositions (Zack, 2014), and operate on this assumption when dealing with members of other racial groups. Others insist that race is socially constructed, rather than natural or biological. Nonetheless, like other social constructs (eg. money), race has profound material effects on our lived experience. "Race is", to quote Charles W. Mills, "a contingently deep reality that structures our particular social universe" (1998, p. 48). Given widespread assumptions, stereotypes, prejudices, or generalizations about Black people, including their lifestyles and genetic predispositions, medical staff and professionals might implicitly or unintentionally be less responsive to the needs of Black patients—a problem not only of implicit bias but of discursive norms (Ayala-López, 2018). Thus, even when the locus of the causal mechanism that explains racial differences in medical expenditures is located with individuals—here: the medical professionals—the properties that fuel this mechanism are structural: assumptions, habits of thought, stereotypes, prejudices, discursive norms, or generalizations about Black patients.

Yet the shortcomings of the medical system cannot be solely attributed to the behavior of medical professionals. A second set of hypotheses looks at material features. In many parts of the US, White patients are geographically closer to medical resources than Black patients. As

---

[6] See Appiah (1990) for an influential analysis of racism. Roughly speaking, Appiah defines "intrinsic racism" as the belief that a racial group is intrinsically superior or inferior to others.

Probst et al. put it, "[d]isadvantage among rural racial/ethnic minorities is a function of place as well as race" (2004). This structural feature—geographic location, affordance, and travel costs—can, in part, explain the racial differences in medical expenditures that led to the racial bias in the health AI.

A third set of hypotheses turns to the patients. Black Americans trust the medical system less than other groups (Boulware et al., 2003). This lack of trust can mediate a lack of engagement in care (Eaton et al., 2015), which in turn leads to lower average medical expenditures. The root causes of this lack of trust are largely unclear. Often, researchers seek to explain this lack of trust by pointing to the infamous Tuskegee study. This would be a rather non-structural explanation. However, the Tuskegee study is not sufficiently well-known in the Black American population to directly explain the lack of trust in this way (Brandon et al., 2005). Lack of trust, even if not necessarily a structural feature itself, is likely best explained in reference to structural features—prevailing narratives, expectations, stereotypes about the medical system. Thus, lack of trust is another structural feature for racial differences in medical expenditures despite identical health conditions.

Despite their variety, these hypotheses still are not exhaustive. Other explanations involve socioeconomic status, gender norms (especially traditional masculine norms that eschew physical vulnerability), lack of awareness of healthcare needs, religious and spiritual attitudes towards medicine, as well as criminal background (Cheatham et al., 2008). Many of these hypotheses point in the direction of further structural explanations.

This example of AI in healthcare illustrates two key points: First, structural explanations are common in the social sciences. The explanations offered for why Black patients have lower medical expenses are a case in point. How medical professionals respond to Black patients, the geographic distance to hospitals, and the lack of trust in the medical system are components of an underlying social structure. It is such structural features that explain differences in average medical expenditure. Second, not surprisingly then, structural explanations are indispensable when you seek to understand, and anticipate, how AI begets injustice. In the case of the healthcare algorithm, the injustice—not just any old "bias"—crept into production in that the feature of "health" was operationalized as medical expenditures.

## 2.2 The Theory of Structural Injustice

One of the most influential and detailed accounts of structural injustice has been articulated by the philosopher Iris Marion Young. The *structure* of society, as Young characterizes it, is "the confluence of institutional rules and interactive routines, mobilization of resources, as well as physical structures such as buildings and roads" (2006). Notice how the structure includes not only rules and conventions—such as implicit assumptions about Black people—but also material properties—the distance to the nearest emergency room. At the same time, *informal social norms and practices* that are not governed by formal rules and conventions, such as stereotypic beliefs, are important constituents of what we mean by "social structure".

Young (2011, pp. 43–45) offers an instructive example of structural injustice, the hypothetical case of a single mother, Sandy. Sandy faces eviction because her apartment building was bought by a property developer. When looking for a new home, Sandy realizes that she is unable to afford most apartments that are geographically close to her workplace. After some deliberation, Sandy decides to rent an apartment 45 minutes away from her workplace, and reduce the length of her commute by also buying a car that she can drive to work. Unfortunately, Sandy's story has a tragic ending: before Sandy is allowed to move in, her prospective landlord requires that she cough up three months' advance rent. Having spent her savings on the new car, Sandy simply doesn't have the money. She and her children now face the terrifying prospect of homelessness.

Fictional as the case may be, it is no stretch to say that many people have found themselves in similar situations to Sandy's. The case is stylized but sufficiently realistic—perhaps even typical. Two things are noteworthy about this case.

Firstly, the case illustrates how unjust outcomes are brought about by agents with good intentions.[7] The injustice that Sandy endures—being evicted and rendered homeless despite her attempts to make the best of the situation—does not result from the actions and decisions of agents who are out to get her. It is perfectly imaginable that the other characters in this story, such as the landlords, bank employees, mortgage brokers, and property developers, have treated Sandy with decency and respect. They may even be *doing their best* as they face personal struggles of their own. Sandy's original landlord may have decided to sell the building because his financial situation made  it impossible for him to maintain it to the standards he should (Young 2006). The point of Young's story is instead that agents causally contribute to Sandy's homelessness *because* they were acting within the law and according to widely accepted norms and moral rules.[8]

Secondly, identifying any individual action or policy that is wrongful or that causes all that is ethically problematic about this situation seems hard or even impossible. Structural features and not individual actions ultimately explains the predicament that Sandy finds herself in. Sandy's case hence illustrates on an individual level what we have seen in the aggregate in the medical case. As in the case of Sandy, the fact that Black Americans have overall lower medical expenses is not explained by individual choices or policies. Structural injustice stems from many hands and many circumstances. Naturally, material possibilities, stereotypes, geographies, and social norms are not the *only* drivers of racial disadvantage, but they hold much and important explanatory value. The role of agents who are simply "following the rules" or "doing as expected" cannot be overstated.

---

[7] Again, we take for granted here the evaluation that would be given by a theory of justice—that Sandy's situation is an injustice.

[8] The formulation here should draw attention to the methodological challenges of structural explanations: Although individual agents causally contribute to the outcome or phenomenon, the structural features are an essential part of the explanation nonetheless.

The case of Sandy hence illustrates the judgment that we started with: No individual acts in a vacuum. The idea of a social structure fills out the situated and context-specific spaces in which we as individuals may often take ourselves to be acting. Accordingly, structural explanations have considerable explanatory power, and are a staple type of explanation in the social sciences (Little, 1991, Chapter 5; Haslanger, 2015; Soon, 2021). Structural explanations can explain how Sandy ended up facing homelessness despite no individual wrongdoing and, perhaps even, everyone's best intentions. Structural explanations form the basis for theories of structural injustice.

The concept of *structural injustice*, as we understand it, brings the power of structural explanations to normative analysis. Structural injustice leverages structural explanations and combines them with a theory of justice to enrich analysis, reasoning, and responses to injustice. Three aspects are noteworthy.

First, structural injustice shifts the focus of our normative attention. It starts by seeking to understand injustice instead of theorizing justice. This is a change of focus in both normative as well as empirical theorizing. Normatively, the approach of structural injustice aims to provide a positive account of injustice, that is, an evaluative theory that conceptualizes injustice not just as the absence of justice or the distance to some ideal of justice. Empirically—our focus in this chapter—structural injustice builds on structural explanations. Structural explanations afford structural injustice great explanatory power. Structural explanations explain why particular nameable groups are persistently disadvantaged. Moreover, structural explanations are unifying. Structural explanations bring into focus commonalities between otherwise disparate-seeming phenomena—the structural features that show up among both causes and consequences. Structural injustice is hence a more holistic or fundamental normative evaluation. This shift in focus—on injustice and on structures—is a hallmark of structural injustice.

Take the example of a police officer who defends herself with objectionable force against a Black man—much like how Amber Guyger fatally shot Botham Jean in his own apartment. Empirically, structural injustice tries to explain how the shooting came about. This violence arises not just from individual behavior but from social norms and practices, and our human tendency to reproduce and reinforce them—by working together, upholding rules and conventions. The social structure constrains and drives individual actions. The structure that underlies police violence constitutes the criminal justice system, the operation of police departments, and negative cultural stereotypes about Black persons. Structural injustice thus identifies these as the targets of reform.

Second, structural injustice is forward-looking. It focuses on structural features. Yet, that does not mean that structural injustice seeks to rectify historic injustice—although it could be complemented in this way. Instead, structural injustice distinguishes between triggering and maintaining causes: The relevant structural causes of racial disadvantage today differ from the past structures that *initially brought about* Black disadvantage. Racial disadvantage may remain alive and well even when state-sanctioned racial discrimination has come to an end (Nuti, 2019). So in a way, the structural injustice has a temporal dimension and has important historic origins. However, decades away from slavery and the Jim Crow era, racism operates now through novel structrual mechanisms that past structures created, such as poverty, new forms of oppression, and different social norms. Structural injustice seeks to understand and reform the structural causes of injustice that operate today.

Finally, structural injustice accumulates. Compared to police violence or Sandy and her family becoming homeless, the example of AI in healthcare appears less troubling. Yet, this impression is misleading. On its own, this disadvantage caused by the healthcare algorithm may seem relatively "small". It may not necessarily make a meaningful difference to health outcomes given that some patients can advocate for themselves and given that there are alternative treatment options. Yet small disadvantages add up over time. Taken together, they explain why the average health of the Black population is lower.[9] Moreover, being less likely to receive preventative healthcare treatment will have negative knock-on effects that lead to further disadvantage. In this way, structural injustice *accumulates* or *compounds* seemingly small disadvantages that derive from agents' tendency to comply with, or operate within the constraints of the status quo.

In sum, the approach of structural injustice has three key features. First, it builds on structural explanations. That is, structural injustice explains phenomena not with reference to individual "micro" attributes (actions of individuals or collective agents) but to broader "macro" attributes (widespread habits of thought, commonplace social practices, compliance with formal or informal norms). Second, structural injustice is forward-looking. It aims to identify and explain the maintaining causes of injustices in order to reform them. Third, injustice and disadvantage accumulate. On their own, an individual injustice might be trivial. The significance of structural injustice can be properly appreciated only when looking at the big picture, where individuals must simultaneously contend with *many* types of disadvantage and the constraints they collectively impose. As Young has stated, "The *accumulated effects* [our emphasis] of past decisions and actions have left their mark on the physical world", in a way that forecloses future possibilities" (2011, p. 53).

## 2.2 Structural Injustice Governs AI

We can now apply the theory of structural injustice to AI. Going beyond the example of AI in healthcare, structural features influence the development and deployment of AI at all steps along the AI pipeline (see the chapter on fairness in this volume for examples). Let us highlight some of these steps.

For starters, structural features influence research agendas, methods, and the choice of problems to tackle. AI research is highly resource-intensive and very expensive. For this reason, powerful economic actors typically decide what problems to tackle and how. Structural features—economic power, political and cultural influence—in part explain which AI is developed and deployed. Looking at who funds and directs AI research institutes that investigate the "ethics" and "fairness" of AI, even at universities that purport to uphold academic freedom and not shy away from critique, you will find it hard to resist the impression that the fox is guarding the hen house (Le Bui & Noble, 2020). Indeed, the fox can just buy the hen house, or, in fact, the whole farm. Next to regulatory capture and cultural capture (Kwak, 2014), in AI governance there is now the problem of academic capture. Similarly, on a smaller scale, individual researchers or public administrators play a causal role in the governance of AI. Again, structural features relating to the social identities of AI's primary movers and shakers—being white, male, having a certain class background—in part explain how AI is developed and used. The structural lens thus brings into focus a strategic analysis of capital interests and ideology, and the causal relevance of social and economic categorical differences between individuals. Combine this structural explanation with a theory of justice, and you may get the result: AI is a form or a tool of structural injustice.

---

[9] The causal story is, of course, more complex. Health outcomes depend not only on material and economic but also on social factors.

Second, structural injustice is reflected in the data. The case study on AI in health provides an example: Patients who are similarly healthy differ in the medical expenses they incur, depending on their race. Similarly, crime data reflect policing practices just as they reflect actual criminality. In short, social structures explain the patterns of behavior and phenomena that data "represent", and social structures condition practices that generate these data. In the case of Sandy, the available data might fail to account for her plight and the complexity of her case, but Sandy's story is likely to show up in data as in the form of significant disparities between different social groups: geographic segregation by race and class or the intergenerational transmission of wealth and opportunity. In an unjust status quo, data evidences—or can even be a driver of—structural injustice.

Third, social structures shape the understanding and meaning of target variables (Fazelpour & Danks, 2021; Passi & Barocas, 2019). Consider labels such as "gender", "race", "health", "criminality", "creditworthiness", or "academic potential". These target variables do not merely represent  things that are "out there" in the world within a model. Instead, such labels operationalize, encode, or calcify social concepts that are in flux. This is not just a semiotic matter. The possible values of the "gender" variable imply a certain substantive view of what gender is—is it social or biological, a binary or a continuum? Moreover, the meaning of "gender" can significantly change the results of causal analyses (Hu & Kohler-Hausmann, 2020). The analytical relevance and the material effects of the choice of data labels make data labels a matter of structural injustice.

The following two points relate to fairness. There is a consensus in AI governance that AI should be fair. However, this focus on fairness is limiting in important ways. The theory of structural injustice makes clear why pursuing fairness is not enough.

Fourth, structural injustice contributes heavily towards epistemic limits in determining whether an individual was fairly treated.  epistemology of fairness. Fairness requires treating like cases alike, but structural injustice makes it hard to tell which cases are alike in the first place.Was Lakisha not hired because of her race, or because Emily was objectively more qualified? If race played a role, then the decision to hire Emily was unfair: Lakisha and Emily are alike (in relevant respects) but were not treated alike. Similarly, did Sandy have to cough up a large deposit for her new apartment because of stereotypes about the financial responsibility of black mothers? Fairness says that differences should matter to the degree that they exist in a just society. However, in a society rife with structural injustice, it is hard to distinguish between those differences that are caused by injustice and those differences that would persist even if the society were just.[10] This larger epistemic problem for fairness is compounded by a smaller one, namely, the fact that structural injustices accumulate and are therefore hard to track. In sum, structural injustice makes it epistemically hard to be fair (Zimmermann & Lee-Stronach, forthcoming). Any AI that aims to be fair hence needs to account for structural injustice [Kate's recommendation]. In a slogan, there can be no fairness without an understanding of social structure and what injustice is due to structural maintaining causes. Even as AI might offer new opportunities to formalize and account for structural difference and injustice (Herington, 2020; Kusner & Loftus, 2020), epistemic limitations remain (Ludwig & Mullainathan, 2021).

Fifth, entrenched social structures limit the efficacy of fairness for justice. Fairness often fails to produce justice, similar to how equality fails to produce equity. "Fairness", like "equality" is often understood as a formal condition or an intrinsic virtue of a decision procedure—think of how the maxim to treat like cases alike is a potent source of disparate treatment, under which persons of marginalized social identities are disadvantaged, intentionally or not, because of the failure to recognize salient differences between various groups. In theory, it may be "fair" for prestigious degree programs to only admit students who score high on standardized tests, insofar as "like" candidates are accepted or rejected on the basis of a criterion that applies to all prospective students. All the same, such requirements have had a disparate impact on members of communities who, owing to structural injustices, have lacked the educational resources to score relatively well on standardized tests (rather than being inherently less competent or suitable fits for the university's program). In an unjust status quo, in which injustice is maintained by social structures, a focus on fairness makes it instrumentally hard or perhaps even impossible to promote justice. Justice or "equity", by contrast, may license *unfair* treatment for reasons of justice (Vredenburgh in this volume). Think, here, of preferential hiring and affirmative action. Such measures can affect structural change—by changing stereotypes, enabling role modeling, affording recognition—but such measures are arguably unfair in this specific sense. Social structure explains why a focus on fairness in AI might be insufficient for promoting justice.

Finally, social structures affect how AI interacts with the context in which it is deployed (consider cases of disparate impact). The lens of social structural explanations is indispensable for anticipating and analyzing the impact of AI. For example, if automatic license plate readers are deployed among arterial roads where Black Americans are more likely to live, Black Americans are more likely to be subject to the negative consequences of being falsely matched with a license plate on a "hot list" of wanted plates. However, not all effects of structural injustice are so easy to anticipate. Most are not. For example, economic theories of home mortgages hide and entrench structural injustice (Herzog, 2017). If even social scientists struggle to capture structural injustice, engineering program managers, public administrators, or computer scientists cannot hardly be expected to succeed on their own. This is an important governance problem because understanding structural injustice is crucial for anyone seeking to anticipate and analyze the impacts of AI.

---

[10] Audit studies are one way of identifying such differences. The "Lakisha" in this paragraph is a reference to one prominent audit study (Bertrand & Mullainathan, 2004). However, their methodology is controversial.

# 3. Existing Normative AI Governance Frameworks

We have defended the structural injustice approach to the governance of AI. Other approaches are available. The example of the medical AI could also be analyzed in terms of harms and benefits—or in terms of values such as DEI. One could say: The algorithm harmed Black patients. One could also say: The algorithm violated the value of equity, the training data was not really diverse, and the development process not inclusive. Given the existence of such alternative approaches, why choose the approach of structural injustice?

This question is particularly pressing because structural injustice can be hard to grasp—harder than harms and benefits, at the very least. Structural injustice raises formidable methodological challenges for researchers and policy-makers who want to draw on this theory.[11] Given all this, is the approach of structural injustice worth it? We argue that structural injustice is indispensable in the conceptual toolkit. The theory of structural injustice has substantial advantages over other ways of approaching normative and social problems of AI.

## 3.1 Harms and Benefits

One framework for analyzing the effects of AI builds on the concepts of harms and benefits. This approach is intuitively compelling and familiar from the ethics of medical research. A particular intervention or artifact—in this case, the use of AI—is evaluated by considering whether there is a risk that individuals would be worse off than they would be without the intervention. Does AI pose threats to their well-being, to their opportunities, or to their health? Are these harms outweighed by benefits?

Such a harms-and-benefits approach surely has its place, but it suffers from severe potential limitations.[12] First, this approach fails to capture ethical problems in full. For starters, it typically restricted to analyzing *individual* harms and benefits. It may have a harder time attending to *collective* harms. To go back to an earlier example: Marketing historically Black neighborhoods in NYC as dangerous, exciting, and exotic does not seem to harm any individual in particular. In theory, it could benefit a Black property-owner who leases out his apartment on

---

[11] Here, we refer to a  challenge that we cannot pursue in this chapter. It divides up into two sets of issues. First, what is the relevant knowledge required to grasp structural injustice? Could the knowledge be conveyed by social-scientific theories? If so, which ones? Alternatively, is such knowledge impossible to quantify or formalize? Second, and relatedly, how can this knowledge be obtained? Do some individuals, because of their social position or role, have better access to the relevant knowledge?

[12] A good example of a harms-based framework in AI that avoids some of these problems is Microsoft's Azure Architecture Application Guide  (Microsoft, 2021).

Airbnb. Nevertheless, even language that is not obviously racialized may affect stereotypes and norms about Black Americans, in a way that is not best described as a "harm" let alone one that is separately identifiable and affects specific individuals. Similarly, the use of beauty standards in advertising might not be harmful at all, let alone be harmful to a *specific* woman, but it may promote distorted beliefs about women as a whole and be a form of structural injustice (Widdows, 2021). Moreover, remember how structural injustice accumulates and compounds. Whereas an approach of harms and benefits identifies individual harms, it may fail to see the fuller picture of how these harms relate. The approach of structural injustice, by contrast, sees disadvantages holistically, compounded by others and adding to other disadvantages in turn.

Second, the concepts of harms and benefits restrict the scope of ethical aspirations and values. Many values are not reducible to harms and benefits.[13] For example, some see it as important that an AI system is explicable or accountable to those subject to it (for clarification on what this means, see other chapters in this section). But the lack of explanations or the absence of accountability does not necessarily constitute a harm. Moreover, some technologies could be beneficial for individuals but still be morally wrong. For example, one can benefit from an intervention that they *have not* consented to. Suppose that, with the help of AI, your employer (or partner) secretly tracks your daily activities, including your dietary and exercise routines. They use these data to serve you lunches that optimize your health and well-being. Despite this benefit, intrusive surveillance without consent is morally off-putting, disturbing, and perhaps morally impermissible. In sum, again a framework of harms and benefits fails to capture the full picture. Governing AI with an eye only to harms and benefits would hence be a mistake. The approach of structural injustice, by contrast, brings into focus structural features such as class interests, economic power, or oppression—concepts that cannot be analyzed purely in terms of harms and their combination.

A third problem with a harms and benefits approach is that a workable account of harms and benefits will need to be accompanied with a theory of how, exactly, harms and benefits ought to be weighed and aggregated. Suppose that some unfortunate individual, Jones, has suffered an accident in the transmission room of a TV station (Scanlon, 1998, p. 235). Jones could be saved from one hour of excruciating pain, but to do so, we would have to cancel the broadcast of a football game, interrupting the pleasure experienced by enthusiastic football fans who are excited about the game. Intuitively, the harm that Jones suffers outweighs the harms that football fans would experience as a result of the canceled transmission. But this judgment depends on a theory of value—and a controversial one. One might argue that, if the number of football fans was sizable enough (e.g. millions of viewers), their collective pleasure might outweigh Jones's suffering. The aggregate benefit is greater than Jones' individual harm. Thus, despite its superficial simplicity, a harms and benefits approach requires a deeper set of principles that describe the aggregation of harms and benefits over individuals and over time. In the case of AI, we need a similar set of principles to justify why certain benefits (e.g. efficiency or economic benefits) ought to be outweighed by considerations of racial justice. But such a set of deeper principles is likely contested and incompatible with the value pluralism—and valuable

---

[13] Here, we understand "harm" as the setting back of one's interests.

pluralism—in societies. The harms and benefits approach is thus neither theoretically simple and, likely, often incompatible with pluralism.

## 3.2 Values and Principles

Another approach is that of value statements or ethics principles. Computer scientists, data scientists, or AI practitioners might have to take a pledge on some values or code of conduct, similar to the Hippocratic Oath (e.g. O'Neil, 2016, Chapter 11). Organizations might articulate their values, codify them, and bring them to life in their organizational culture and processes. For a while, technology companies, public bodies, and professional associations prolifically listed their principles and values—they may say that technology should be used non-discriminatorily, and that AI should be explicable. *DEI is, in part, an instance of this approach.* An organization may say that they are committed to diversity, equity, and inclusion just as they may say that they are committed to explainable AI—with all the good that this entails: The organization will have processes to determine the meaning of "diversity", "equity", and "explainability" and to make sure its conduct is informed by these values. Moreover, when an organization has publicly committed to such values, it can be held to them, from within as well as from without.

Although such codes of conduct have their place (Davis, 1991; Stevens, 2008), similar to ethical analyses based on harms and benefits, their efficacy is limited (Mittelstadt, 2019; Whittlestone et al., 2019). Statements of values—even if they are articulated in detail, and even if these values are sincerely held and underwritten by a public commitment and organizational structures—are not a viable general approach to AI governance. DEI extends the list of organizational values and principles but suffers from the same shortcomings of this more general approach. There are risks of window-dressing, ethics-washing, and cheap talk. Organizational values might become static and ethical governance might turn into a compliance exercise of box ticking (Boddington, 2020). More importantly, statements of values or principles are not a viable ethical framework for the governance of AI for three reasons.

First, some organizations may see AI ethics as their mission but cannot, on their own, bring values to life. Examples here are professional organizations, such as the Association of Computing Machinery (ACM), the American Statistical Association, or the American Society for Public Administration. Such organizations lack the processes to explicate, role-model, incentivise, or enforce the values that they give themselves— processes that are crucial for accountability and for codes to be effective in a governance context (Gasser & Schmitt, 2020; Stevens, 2008).

Second, the task of governance may involve many actors with diverging or competing values. The approach of articulating ethical values and principles, and bringing them to life, cannot do much to reconcile differences. Similar to the problem of the harms and benefits approach, this is a major shortcoming since value pluralism is central to many issues—from the ethics of autonomous vehicles to the value alignment problem (Gabriel, 2020; Himmelreich, 2020).

Of course, such pluralism could be accommodated, if the principles are limited to some basic consensus. Such consensus or minimal principles could be ethical guardrails or democratic principles and values. In this vein, the state—and especially its courts and executive agencies—often purport to be based on values of this sort, neutral values or consensus values. But whether such neutrality is feasible and whether it is desirable is questionable (Wall, 2021, sec. 3.1). Moreover, there is likely a tradeoff between a set of values that is neutral and that can find consensus on the one hand, and a set of values that is interesting, promotes justice, and guides actions. Principles and values that may pass muster and count as neutral enough—such as respect for value pluralism or freedom of speech—might just not be informative enough to guide actions or regulations, let alone promote justice (Himmelreich, 2022).

## 3.3 Justice

A third possible evaluative framework centers around the idea of *justice.* Theories of justice broaden our understanding of the *unit of evaluation*—beyond harms and benefits—and they account for the *foundations of values*—beyond merely listing or stating them. Theories of justice aim to orient reasoning and discussions of regulation and policy—and in this sense inform policy-making—while they aim at the same time to have a broad appeal. This is because theories of justice are meant to regulate issues of *common* concern. The idea is to have a theory to regulate how we get along, in a way that makes room for conflicting views and beliefs about what is morally right and good. Theories of justice start from the idea that there are moral conflicts and deep disagreements on matters of common concern. All this makes theories of justice the most useful and fitting lens to analyze normative issues of AI.

In both ordinary and philosophical discourse, the words "ethics" and "justice" are frequently run together. It is commonly assumed that what is *ethical* is also *just*: the measure of justice, in a particular society, is how ethically the state treats its citizens and other persons who are subject to its power. Politics, then, is seen as amounting to applied ethics. Pressing political questions—eg. whether capital punishment is morally permissible, or whether a state may block immigrants from entry—are ethical questions similar to, say, whether eating meat or buying fast fashion is morally permissible.

Justice, by contrast, is concerned with normative requirements or considerations that are different—or have different emphasis—from those of ethics more broadly. As Bernard Williams writes, political philosophy should "use distinctively political concepts, such as power, and its normative relative, legitimation" (2005, p. 77).[14] According to Williams, the structures we live within must *make sense* to us, to the extent that we are able to see why it would be wrong for us to reject or resist those structures. For theorists like Williams who strongly distinguish between "ethics" and "politics", the central puzzle of politics is not "are we being treated ethically?", but rather, "can we *make sense* of the exercises of power that we are routinely subject to?" Justice hence is not a matter of compliance, of applying principles, or of living by values but it is instead decidedly practical and involves processes, such as public deliberation and contestation. Put differently, justice is *not* primarily about treating people in line with *ethical* principles: instead, for justice, exercises of power must be *justified* to those subject to them. The approach of justice brings into focus questions of who may issue rules and whose word counts (authority), the processes in which such rules are made and enforced (legitimacy), and the reasons for the rules, decisions or actions (justification).

---

[14] We use Williams here as an illustrative slogan since we disagree with the meta-normative view—political realism—that this quote conveys.

The task of theories of justice is to normatively ground the regulation and interrogation of power. This applies to the legal system as well as to the markets and the economic system. Such systems cannot take *any* form that power-holders desire. They must be built or maintained in a way that *makes sense* to the persons who live within them. This is because, as Rawls insists, power must be legitimated to us because of its "profound and present" effects on our lives—our life-prospects, goals, attitudes, relationships, characters (Rawls, 1971). Suffice to say, a society that persistently disadvantaged or subordinated persons on the basis of gender or race would be extremely difficult, if not entirely impossible, to legitimate. It is more than reasonable, for a Black man who is disproportionately subject to state-sanctioned police violence, to ask *why* he should be required to accept these social conditions. Quite obviously, a racist social arrangement would not "make sense" to him or others in his position. Either way, then, theories of justice—whether as an extension or a reform of existing theories—are the right kind of theoretical framework to address fundamental normative issues in the governance of AI.

Theories of justice spell out ideas of equality, freedom, and community. Such theories explain why racism is wrong, why colonialism is wrong, what oppression is, how it can be overcome, what an absence of these wrongs would look like, and why such a state would be desirable, even under conditions of pluralism. Theories of justice also delineate and ground liberties. They explain, for example, when and why citizens have a right to an explanation from courts or administrative bodies.

The approach of justice has hence several advantages over the alternative approaches. In contrast to harms and benefits, it broadens the unit of evaluation. In contrast to the approach of listing values and principles, the justice-based approach aims not at values themselves but the underlying reasons for values—it answers the question of why we need values like explainability, accountability, or equity at all. And in contrast to both alternative approaches, justice puts dilemmas and conflicts *between* individuals front and center. The approach of justice aims to regulate matters of common concern. It presupposes and respects a meaningful degree of conflict and disagreement.

## Diversity, Equity, and Inclusion in a Theory of Justice

The approach of justice hence is more foundational than the alternative two approaches. We hope it can also be useful. To illustrate, consider how the approach of structural injustice in particular recognizes calls for greater Diversity, Equity, and Inclusion (DEI) as demands of justice. Articulated as demands of justice, DEI is not an attempt on the part of marginalized social groups to secure more power, resources, and advantage for themselves, as it is often uncharitably interpreted. Nor is it a mere matter of generosity or beneficence that would help to make the world a morally better place. From the perspective of justice, the values of DEI stand on reasons that should have a hold on everyone regardless of their self-interest. Such reasons of justice weigh more heavily than the reasons to help others in need. Moreover, reasons of justice are important because the institutions in an unjust society often lack legitimacy and authority.

To illustrate how the demands of DEI can be seen as demands of justice, we need to unpack the content of DEI in more detail. In our view, "diversity", "equity", and "inclusion" are separate but closely interrelated ideas. Out of the three, we understand equity as the fundamental one. Not surprisingly, egalitarian theories of justice require (or assume), among other things, some form of gender and racial equity for justice to obtain. Women and persons of color cannot be asked to accept a society that persistently subjects them to disadvantage. In this sense, in trying to make society acceptable to everyone, DEI—and especially equity—formulates a partial ideal of justice.

Moreover, the values of diversity and inclusion are instrumental for, or even constitutive of, achieving this ideal of equity. For example, the tech industry remains heavily male-dominated, and women have been twice as likely to leave as men.[15] Arguably, gender equity can only be achieved if tech corporations aim for gender diversity—for example, by actively recruiting and retaining women. In the absence of women practitioners who are attentive to existing gender-based disparities, there is considerable risk that AI will inadvertently perpetuate structural gender injustice. At the same time, to be truly inclusive towards women, tech corporations must reconsider their professional norms and practices, and how those may be hostile or exclusionary to women. Sexist stereotypes about women's inherent lack of suitability for STEM fields, which can lead to biased and unfair treatment towards women employees, must also be resisted.[16]

Approaching AI ethics through the lens of justice may not just vindicate and give structure to the ideas of DEI. The approach of structural injustice may also enrich and clarify their content. The idea of "diversity" can be—and for the purposes of US constitutional law, often has been—understood as symmetric: A relatively homogenous group of students is technically made more "diverse" by anyone who differs from the group members in one of many possible dimensions, but the justice-based approach will interrogate the reasons for valuing "diversity" and, from there, inform our sense of *which* diversities are relevant—this is why gender and racial diversity have taken center-stage in attempts to "diversify" particular spaces, rather than diversity in properties like hair or eye color.  As we have seen, justice takes into account persistent, and even historic, disadvantages that are connected to these social identities—see, for example, Charles Mills's reminder of the importance of corrective justice (2017, p. 208). To make matters worse, "diversity" is often understood as an individualistic notion: An individual improves a group's diversity simply in virtue of their intrinsic properties. On this view, add an engineer of color to the team, and the work of DEI is done. By contrast, an approach of justice—especially an approach of structural injustice as that of Iris Young (2009)—looks to overall social conditions that determine or constrain our possibilities, not merely to individual contributions to individual groups. It serves as a reminder that structural disparities often continue to obtain even when certain individuals from diverse backgrounds may achieve great success within their occupation.  Existing theories of justice and structural injustice hence may not only ground the ideas of DEI, but may also enrich and clarify their content.

---

[15] See: https://www.theatlantic.com/magazine/archive/2017/04/why-is-silicon-valley-so-awful-to-women/517788/

[16] The pursuit of diversity and inclusion also relates to other ideas and values, not just to justice. Although we focus on diversity and inclusion as requirements of justice, there can be other valuable instrumental benefits to diversity and inclusion, such as greater epistemic performance of a group. For example, with more diverse input, tech corporations might get better at developing helpful algorithms.

# 4. Conclusion

This chapter defends two important parts of a framework for the governance of AI: structure and justice. We argued, first, that an approach to the governance of AI should avail itself of the analytical benefits of structural explanations, and, second, that the evaluative component of such a framework should be provided by a theory of justice. We illustrated the advantages of structural explanations and how an approach of structural injustice recognized and advances the values of DEI.

In conclusion, it is time to look forward. First, we sketch a relevant theoretical question in how theories of justice relate to AI. Second, in a more practical vein, we outline how the approach of structural injustice can be accompanied by an understanding of responsibility.

## 4.1 AI as Part of the Structure?

A theory of AI justice can take two forms. First, existing theories of justice can be applied to AI. This strategy has clear prospects: Theories of justice can explain why AI should be explainable just like other decisions should be explainable. The demand for explainability in AI may fall out of a more general obligation to explain judicial and administrative decisions to those subject to them.

Second, instead of merely applying theories of justice to AI, AI can be the novel subject of a theory of justice. This second theoretical avenue is motivated by the thought that, broadly put, AI changes the nature of society—the social structure, the subject of justice. AI not only augments existing social practices, and AI does not just realize existing social functions in a novel technological way. Instead, the idea is that AI raises questions of justice just like the tax system, the financial system, the criminal justice system, the judicial system, or the social system—e.g. interpersonal interactions and family—raise substantive questions of justice. On this second approach, AI is seen as part of the social structure.

Justice and structure are closely related. A focus on justice requires a focus on structure. This idea goes back to John Rawls whose work is foundational for contemporary theories of justice.[17]

---

[17] Rawls asserts, "the primary subject of justice is the basic structure of society, or more exactly, the way in which the major social institutions distribute fundamental rights and duties and determine the advantages from social cooperation" (1971, our emphasis). While we do not seek to defend a Rawlsian

In Rawls' picture, the subject of justice is the *basic structure* of a society, that is, all institutions, taken together and over time—including political, legal, economic, social, civil systems—that form the backdrop for life in this society.

Rawls—not surprisingly—assumes a narrow and dated idea of the basic structure in several ways. First, Rawls pays relatively little attention to the individuals in the structure (Cohen, 1997). Individuals need to obey the law but they are not subject to demands of justice directly. Rawls concentrates instead largely on "the political constitution and principal economic and social arrangements". However, Rawls's articulation of the basic structure falls short in a different respect. While it briefly includes the "monogamous family" as an example of a "major social institution (Rawls 1971), the feminist philosopher Susan Okin sharply critiques Rawls's inattention to families—including family structures that are not necessarily "monogamous" (Okin , 1989, p. 93) and provides a detailed elaboration on how family units can be major sites of social inequalities. Women are typically assumed to be responsible for performing the lion's share of domestic chores and caregiving labor, which may severely hamper their opportunities relative to men. Additionally, families are the "first schools of moral development, where we first learn to develop a sense of justice" (Okin, 1989, p. 31). Here, the implication is that unjust family structures can hamper or distort our nascent capacities to observe just terms of cooperation under circumstances of conflicting interests and scarce resources. For these reasons, Okin believes that families must also be regulated by principles of justice. It is best to think of the basic structure as including not only the legal system and the economy, but also the family. The idea of the basic structure is thus not only closely related to ideas of justice, what counts as the basic structure is also not set in stone, not even when we think about justice in the tradition of Rawls.

We take Rawls's willingness to adapt his definition of the "basic structure", in the face of societal upheaval, to be a sign that the idea of justice can *and* should be responsive to the growing ubiquity of AI.

In the age of AI, and the overwhelming power of "Big Tech" to shape our everyday lives, AI might have to be seen as an important component of this "basic structure". The question is thus: Should AI be seen as part of the basic structure?

Iason Gabriel (2022) argues that it should. He argues that the basic structure of society is best understood as "a composite of *socio-technical systems*" that bring about "new forms of stable institutional practice and behavior". It would be mistaken to think of "the political constitution and principal economic and social arrangements" as somehow removed or insulated from its interactions with technology. Instead, AI "increasingly *shapes* elements of the basic structure in relevant ways". It *mediates* the distribution of basic rights and duties, along with the advantages of social cooperation (Gabriel, 2022). The healthcare algorithm above is only one example. Public services—from policing to welfare, housing, and infrastructure—are increasingly theory of justice, nor is it necessary to spell out the exact principles of justice that Rawls has proposed, his work is foundational to how we understand the point and purpose of a theory of justice.

automated, with profound aggregate effects on those who are already disadvantaged (Eubanks, 2018).

However, the question of how to best develop theories justice for AI—by applying existing theories or by rethinking conceptions of the basic structure—is still open. It is a crucial question for ongoing normative and empirical research.

## 4.2 AI Justice via Responsibility

In closing, we want to highlight one practical upshot of structural injustice. Structural injustice emphasizes individual responsibility (Goodin & Barry, 2021; McKeown, 2021). The cultivation of practices of responsibility is thus an important topic for AI governance.

Injustice raises a question of responsibility: Who is responsible for rectifying structural injustice? Two answers may immediately come to mind.

The first answer is that *everyone* is responsible together. Institutionally, this means that the state might be responsible. States are already responsible for discharging duties of justice to their citizens. States implement ideas of distributive justice: they tax, subsidize, incentivise in order to allocate resources and opportunities. Private entities, by contrast, are not bound by considerations of justice. Thus, they are not responsible for rectifying structural injustice.

The second answer is that *nobody* is responsible. After all, structural injustice emphasizes that agents contribute to structural injustice despite the best of their intentions, that is, without wronging anyone. It may thus be inappropriate to hold individuals responsible, let alone blame them for structural injustice.

Both these answers are unsatisfactory.[18] That states alone are responsible for rectifying structural injustice is unconvincing. Our chapter has emphasized the growing power and profound effects that AI has on individuals' life-prospects. A state-centric view of political responsibility would betray an overly limited view of the social structure. Many non-state actors—"Big Tech" and billionaires—shape AI, whether AI itself is part of the basic structure or not.

Moreover, both answers assume an overly narrow view of responsibility. They falsely assume that responsibility entails blameworthiness. Instead, we can distinguish between *attributive* and *substantive responsibility.* Attributive responsibility "helps us decide to whom we should attribute (retrospectively) praise or blame for a particular state of affairs" (Parrish, 2009). Substantive responsibility, by contrast, refers to what people are required to do (Scanlon, 1998, Chapter 6).

Suppose that, in a fit of anger, I spitefully pour water onto your laptop to damage it. Here, I am clearly *attributively responsible* for destroying your laptop; it is appropriate for you to blame me for what I have done. Of course, I am also *substantively responsible* for paying for your laptop. By contrast, suppose instead that I have accidentally spilled water on your laptop. It was an accident with no fault on my part. In this scenario, I am *not* attributively responsible for damaging your laptop but I can still be substantively responsible, due to the causal role I have played in damaging the laptop. Paying for the repair is still on me.

---

[18] Given the purposes and limitations of this chapter, our treatment of the relevant substance here will be very superficial.

For structural injustice, it seems best to focus on substantive responsibility. Even if individuals are not to blame for structural injustice, they can still be responsible for rectifying it. The goal of responsibility here is to identify and address individual actions, which may be blameless, but that may have generated injustice. Moreover, we can ask what social changes will prevent—or at least reduce the likelihood of—future structural disadvantage, even if they did not causally contribute to existing structural injustice. These are two paths of assigning responsibility, without blame, that operationalize the theory of structural injustice to affect structural reform.[19]

In conclusion, the overall picture that we hoped to sketch in this chapter is this: Structural injustice offers an analytical framework and a normative framework—via structural explanations and a theory of justice respectively. Both of these are, or so we have argued, indispensable for a normative theory in the governance of AI. Moreover, structural injustice offers a useful practical approach by aiming to identify responsibilities of individuals to reform where they can and to rectify without blame.

# References

Appiah, K. A. (1990). Racisms. In D. T. Goldberg (Ed.), *Anatomy of Racism* (pp. 3–17).

University of Minnesota Press.

Ayala-López, S. (2018). A Structural Explanation of Injustice in Conversations: It's about Norms.

*Pacific Philosophical Quarterly*, *99*(4), 726–748. https://doi.org/10.1111/papq.12244

Bertrand, M., & Mullainathan, S. (2004). Are Emily and Greg More Employable than Lakisha

and Jamal? A Field Experiment on Labor Market Discrimination. *The American*

*Economic Review*, *94*(4), 991–1013.

Boddington, P. (2020). Normative Modes: Codes and Standards. In M. D. Dubber, F. Pasquale,

& S. Das (Eds.), *The Oxford Handbook of Ethics of AI* (pp. 123–140). Oxford University

---

[19] To be clear, this does not mean that everyone is equally responsible for rectifying structural injustice. Depending on their social position, some participants in the relevant structure may have made greater causal contributions than others, and consequently have more burdensome responsibilities. Some participants may also have more power than others to rectify structural injustice. Finally, some agents may also still have attributive responsibility and be blameworthy for the injustices in which they are involved.

Press. https://doi.org/10.1093/oxfordhb/9780190067397.013.7

Boulware, L. E., Cooper, L. A., Ratner, L. E., LaVeist, T. A., & Powe, N. R. (2003). Race and Trust in the Health Care System. *Public Health Reports*, *118*(4), 358–365. https://doi.org/10.1093/phr/118.4.358

Brandon, D. T., Isaac, L. A., & LaVeist, T. A. (2005). The legacy of Tuskegee and trust in medical care: Is Tuskegee responsible for race differences in mistrust of medical care? *Journal of the National Medical Association*, *97*(7), 951–956.

Cheatham, C. T., Barksdale, D. J., & Rodgers, S. G. (2008). Barriers to health care and health-seeking behaviors faced by Black men. *Journal of the American Academy of Nurse Practitioners*, *20*(11), 555–562. https://doi.org/10.1111/j.1745-7599.2008.00359.x

Cohen, G. A. (1997). Where the Action is: On the Site of Distributive Justice. *Philosophy & Public Affairs*, *26*(1), 3–30.

Cuevas, A. G., O'Brien, K., & Saha, S. (2016). African American experiences in healthcare: "I always feel like I'm getting skipped over." *Health Psychology*, *35*(9), 987–995. https://doi.org/10.1037/hea0000368

Davis, M. (1991). Thinking Like an Engineer: The Place of a Code of Ethics in the Practice of a Profession. *Philosophy & Public Affairs*, *20*(2), 150–167.

Eaton, L. A., Driffin, D. D., Kegler, C., Smith, H., Conway-Washington, C., White, D., & Cherry, C. (2015). The Role of Stigma and Medical Mistrust in the Routine Health Care Engagement of Black Men Who Have Sex With Men. *American Journal of Public Health*, *105*(2), e75–e82. https://doi.org/10.2105/AJPH.2014.302322

Eubanks, V. (2018). *Automating Inequality: How High-Tech Tools Profile, Police, and Punish the Poor*. St. Martin's Press.

Fazelpour, S., & Danks, D. (2021). Algorithmic bias: Senses, sources, solutions. *Philosophy Compass*, *16*(8), e12760. https://doi.org/10.1111/phc3.12760

Gabriel, I. (2020). Artificial Intelligence, Values, and Alignment. *Minds and Machines*, *30*(3),

411–437. https://doi.org/10.1007/s11023-020-09539-2

Gabriel, I. (2022). Towards a Theory of Justice for Artificial Intelligence. *Dædalus*, *151*(2), 218–231.

Gasser, U., & Schmitt, C. (2020). The Role of Professional Norms in the Governance of Artificial Intelligence. In M. D. Dubber, F. Pasquale, & S. Das (Eds.), *The Oxford Handbook of Ethics of AI* (pp. 140–159). Oxford University Press. https://doi.org/10.1093/oxfordhb/9780190067397.013.8

Goodin, R. E., & Barry, C. (2021). Responsibility for structural injustice: A third thought. *Politics, Philosophy & Economics*, *20*(4), 339–356. https://doi.org/10.1177/1470594X211027257

Haslanger, S. (2015). What is a (social) structural explanation? *Philosophical Studies*, *173*(1), 113–130. https://doi.org/10.1007/s11098-014-0434-5

Herington, J. (2020). Measuring Fairness in an Unfair World. *Proceedings of the AAAI/ACM Conference on AI, Ethics, and Society*, 286–292. https://doi.org/10.1145/3375627.3375854

Herzog, L. (2017). What Could Be Wrong with a Mortgage? Private Debt Markets from a Perspective of Structural Injustice. *Journal of Political Philosophy*, *25*(4), 411–434. https://doi.org/10.1111/jopp.12107

Himmelreich, J. (2020). Ethics of Technology Needs More Political Philosophy. *Communications of the ACM*, *63*(1), 33–35. https://doi.org/10.1145/3339905

Himmelreich, J. (2022). Against "Democratizing AI." *AI & SOCIETY*. https://doi.org/10.1007/s00146-021-01357-z

Hu, L., & Kohler-Hausmann, I. (2020). What's sex got to do with machine learning? *Proceedings of the 2020 Conference on Fairness, Accountability, and Transparency*, 513. https://doi.org/10.1145/3351095.3375674

Kusner, M. J., & Loftus, J. R. (2020). The long road to fairer algorithms. *Nature*, *578*(7793), 34–36. https://doi.org/10.1038/d41586-020-00274-3

Kwak, J. (2014). Cultural capture and the financial crisis. In D. Carpenter & D. A. Moss (Eds.),
    *Preventing regulatory capture: Special interest influence and how to limit it* (pp. 79–81).
    Cambridge University Press.

Le Bui, M., & Noble, S. U. (2020). We're Missing a Moral Framework of Justice in Artificial
    Intelligence: On the Limits, Failings, and Ethics of Fairness. In M. D. Dubber, F.
    Pasquale, & S. Das (Eds.), *The Oxford Handbook of Ethics of AI* (pp. 161–179). Oxford
    University Press. https://doi.org/10.1093/oxfordhb/9780190067397.013.9

Little, D. (1991). *Varieties of social explanation: An introduction to the philosophy of social
    science*. Westview Press.

Ludwig, J., & Mullainathan, S. (2021). Fragile Algorithms and Fallible Decision-Makers: Lessons
    from the Justice System. *Journal of Economic Perspectives*, *35*(4), 71–96.
    https://doi.org/10.1257/jep.35.4.71

McKeown, M. (2021). Structural injustice. *Philosophy Compass*, *16*(7), e12757.
    https://doi.org/10.1111/phc3.12757

Microsoft. (2021, November 12). *Types of harm—Azure Application Architecture Guide*.
    https://docs.microsoft.com/en-us/azure/architecture/guide/responsible-innovation/harms-
    modeling/type-of-harm

Mills, C. W. (1998). *Blackness Visible: Essays on Philosophy and Race*. Cornell University
    Press.

Mills, C. W. (2017). *Black Rights/white Wrongs: The Critique of Racial Liberalism*. Oxford
    University Press.

Mittelstadt, B. (2019). Principles alone cannot guarantee ethical AI. *Nature Machine Intelligence*,
    *1*(11), 501–507. https://doi.org/10.1038/s42256-019-0114-4

Nuti, A. (2019). *Injustice and the Reproduction of History: Structural Inequalities, Gender and
    Redress*. Cambridge University Press.

Obermeyer, Z., Powers, B., Vogeli, C., & Mullainathan, S. (2019). Dissecting racial bias in an algorithm used to manage the health of populations. *Science*, *366*(6464), 447–453. https://doi.org/10.1126/science.aax2342

Okin, S. M. (1989). *Justice, gender and the family*. Basic Books.

O'Neil, C. (2016). *Weapons of Math Destruction: How Big Data Increases Inequality and Threatens Democracy*. Crown.

Parrish, J. M. (2009). Collective responsibility and the state. *International Theory*, *1*(01), 119–154. https://doi.org/10.1017/S1752971909000013

Passi, S., & Barocas, S. (2019). Problem Formulation and Fairness. *Proceedings of the Conference on Fairness, Accountability, and Transparency*, 39–48. https://doi.org/10.1145/3287560.3287567

Probst, J. C., Moore, C. G., Glover, S. H., & Samuels, M. E. (2004). Person and Place: The Compounding Effects of Race/Ethnicity and Rurality on Health. *American Journal of Public Health*, *94*(10), 1695–1703.

Rawls, J. (1971). *A Theory of Justice* (1999 Revised edition). Harvard University Press.

Rini, R. (2020). *The Ethics of Microaggression*. Routledge.

Scanlon, T. (1998). *What we owe to each other*. Belknap Press of Harvard University Press.

Soon, V. (2021). Social structural explanation. *Philosophy Compass*, *16*(10), e12782. https://doi.org/10.1111/phc3.12782

Stevens, B. (2008). Corporate Ethical Codes: Effective Instruments For Influencing Behavior. *Journal of Business Ethics*, *78*(4), 601–609. https://doi.org/10.1007/s10551-007-9370-z

Törnberg, P., & Chiappini, L. (2020). Selling black places on Airbnb: Colonial discourse and the marketing of black communities in New York City. *Environment and Planning A: Economy and Space*, *52*(3), 553–572. https://doi.org/10.1177/0308518X19886321

Wall, S. (2021). Perfectionism in Moral and Political Philosophy. In E. N. Zalta (Ed.), *The Stanford Encyclopedia of Philosophy* (Fall 2021). Metaphysics Research Lab, Stanford University. https://plato.stanford.edu/archives/win2017/entries/perfectionism-moral/

Whittlestone, J., Nyrup, R., Alexandrova, A., & Cave, S. (2019). *The Role and Limits of Principles in AI Ethics: Towards a Focus on Tensions*. 7.

Widdows, H. (2021). Structural injustice and the Requirements of Beauty. *Journal of Social Philosophy*, *52*(2), 251–269. https://doi.org/10.1111/josp.12369

Williams, B. (2005). *In the Beginning Was the Deed: Realism and Moralism in Political Argument*. Princeton University Press.

Young, I. M. (2006). Responsibility and Global Justice: A Social Connection Model. *Social Philosophy and Policy*, *23*(1), 102–130. https://doi.org/10.1017/S0265052506060043

Young, I. M. (2009). Structural Injustice and the Politics of Difference. In T. Christiano & J. Christman (Eds.), *Contemporary Debates in Political Philosophy*. John Wiley & Sons. http://onlinelibrary.wiley.com/book/10.1002/9781444310399

Young, I. M. (2011). *Responsibility for Justice*. Oxford University Press, USA.

Zack, N. (2014). The Philosophical Roots of Racial Essentialism and Its Legacy. *Confluence: Journal of World Philosophies*, *1*. https://scholarworks.iu.edu/iupjournals/index.php/confluence/article/view/522

Zimmermann, A., & Lee-Stronach, C. (forthcoming). Proceed with Caution. *Canadian Journal of Philosophy*. https://doi.org/10.1017/can.2021.17